\documentclass{aastex}
\usepackage{emulateapj5}

\newcommand\nudotdotdot{\ifmmode\stackrel{\bf \,...}{\textstyle
\nu}\else$\stackrel{\,...}{\textstyle \nu}$\fi}
\newcommand{\approxlt}{\mbox{$\;^{<}\hspace{-0.24cm}_{\sim}\;$}}

\slugcomment{Accepted by the Astrophysical Journal}

\shorttitle{Radio Polarization of PSR J1119$-$6127}
\shortauthors{Crawford \&\ Keim}

\begin{document}

\title{Radio Polarization of the Young High-Magnetic-Field
Pulsar PSR J1119$-$6127}

\author{Fronefield Crawford and Nathan C. Keim} 
\affil{Department of Physics, Haverford College, Haverford, PA 19041}
\email{fcrawfor@haverford.edu} 

\begin{abstract}
We have investigated the radio polarization properties of PSR
J1119$-$6127, a recently discovered young radio pulsar with a large
magnetic field.  Using pulsar-gated radio imaging data taken at a
center frequency of 2496 MHz with the Australia Telescope Compact
Array, we have determined a rotation measure for the pulsar of $+842
\pm 23$ rad m$^{-2}$. These data, combined with archival polarimetry
data taken at a center frequency of 1366 MHz with the Parkes
telescope, were used to determine the polarization characteristics of
PSR J1119$-$6127 at both frequencies.  The pulsar has a fractional
linear polarization of $\sim$~75\% and $\sim$~55\% at 1366 and 2496
MHz, respectively, and the profile consists of a single, wide
component.  This pulse morphology and high degree of linear
polarization are in agreement with previously noticed trends for young
pulsars (e.g., PSR J1513$-$5908).  A rotating-vector (RV) model fit of
the position angle (PA) of linear polarization over pulse phase using
the Parkes data suggests that the radio emission comes from the
leading edge of a conal beam.  We discuss PSR J1119$-$6127 in the
context of a recent theoretical model of pulsar spin-down which can in
principle be tested with polarization and timing data from this
pulsar. Geometric constraints from the RV fit are currently
insufficient to test this model with statistical significance, but
additional data may allow such a test in the future.
\end{abstract}


\section{Introduction}

Pulsar polarimetry is one of the keys to understanding the process and
geometry of radio emission from pulsars. In the rotating-vector (RV)
model \citep{rc69} the polarization of pulsar radio emission is linked
to the emission geometry in such a way that as the pulsar rotates, the
axis of linear polarization is aligned with the projected direction on
the sky of the pulsar's magnetic dipole axis.  The pulsar's emission
geometry itself may be described by two angles, each measured from the
pulsar's angular momentum vector: the magnetic inclination angle
$\alpha$ is the angle between the spin axis and the magnetic dipole
axis, and the angle $\zeta$ measures the separation between the spin
axis and an observer's line of sight.  Given this geometric
description, the RV model defines the linear polarization position
angle (PA) $\psi$ as a function of pulse phase $\phi$ according to

\begin{equation}
\tan (\psi - \psi_0 ) = {\sin \alpha \sin (\phi - \phi_0 ) \over
\sin \zeta \cos \alpha - \cos \zeta \sin \alpha \cos (\phi - \phi_0 ) }
\end{equation}

\noindent where $\psi_0$ is the PA corresponding to the projected
direction of the pulsar's rotation axis on the sky, and $\phi_0$ is
the pulse phase at which the PA swings most rapidly, corresponding to
the magnetic axis sweeping past the line of sight.  PA is measured
from North to East on the sky, following the usual convention (e.g.,
Everett \& Weisberg 2001). A more observationally useful replacement
for $\zeta$ is the impact parameter $\beta$, defined as the smallest
angle between the magnetic axis and the line of sight as the dipole
rotates, $\zeta - \alpha$ \citep{ew01}. A small value of $|\beta|$
corresponds to a steep PA swing as the magnetic axis sweeps past the
line of sight.

With sufficient coverage over pulse phase, a fit for RV model
parameters $\psi_0$, $\phi_0$, $\alpha$, and $\beta$ may be performed.
Since $\phi_0$ is the phase corresponding to the center of the
magnetic pole, the fit determines the geometry not only of the pulsar
itself, but of the pulsar's regions of radio emission, which can lie
at various positions relative to the pulsar's magnetic axis. Emission
at the magnetic axis is not always present, and emission may not be
symmetrical about the axis, in some cases giving rise to one or more
pulses that lead or trail the beam center \citep{lm88}.

One possible geometric interpretation of the phenomenology of pulsar
polarization profiles is that emission can come from either core or
conal beams. While a core beam is a narrow, solid cone of radio
emission extending outward along the pulsar's magnetic dipole axis, a
conal beam is in the form of a larger, hollow cone that circumscribes
the core beam. In this interpretation, conal emission tends to have a
steeper spectral index and a higher degree of linear polarization than
core emission \citep{lm88}. While \citet{r83} has proposed that core
and conal components arise from differing emission mechanisms,
\citet{lm88} contend that there is a continuous variation in radiation
properties between core and cone, and that all radio beam emission
shares the same mechanism. \citet{lm88} also suggest that emission
beam patterns may be ``patchy,'' such that only one side of an ideal
hollow cone beam might be active.

A model of pulsar spin-down proposed by \citet{m97} connects pulsar
timing with geometry such that its predictions may be tested through
observation.  The model treats the pulsar and its inner magnetosphere
as a single perfectly conducting sphere rotating in a vacuum. The
model thus differs from the standard vacuum-dipole theory of pulsar
spin down \citep{og69} in which the rotating magnetic dipole is
treated as point-like. In the Melatos model, the components of the
electric and magnetic fields exterior to the rotating conducting
sphere are modified according to calculations first performed by
\citet{d55}. The resulting modified electromagnetic torque is also
dependent upon the magnetic inclination angle $\alpha$; the strongest
braking occurs when the magnetic and spin axes are orthogonal (i.e.,
when $\sin \alpha$ is a maximum).

Using the modified torque expression derived from these fields, the
Melatos model predicts the values of the first and second braking
indices $n$ and $m$, which are determined observationally by $n = \nu
\ddot \nu / {\dot \nu}^2$ and $m = \nu^2 \nudotdotdot / {\dot \nu}^3$,
where $\nu$ is the observed pulsar rotation frequency. The model
determines $n$ and $m$ using only three observable parameters: the
period $P = 1/\nu$, the period derivative $\dot P$, and $\alpha$.  In
the point-dipole spin-down model, theory predicts $n = 3$ and $m =
n(2n-1) = 15$ \citep{br88}. However, in the Melatos model, the
decreased braking torque produces values for $n$ and $m$ which are
smaller than these, and there are no free parameters in the model.
The model is thus highly falsifiable if adequate constraints can be
placed on a pulsar's emission geometry.  The model has so far been
applied with some success to the Crab pulsar, PSR B0540$-$69, and PSR
J1513$-$5908 (B1509$-$58).

PSR J1119$-$6127 is a 408-ms radio pulsar that was discovered in
August 1997 using the Parkes 64-m radio telescope during the Parkes
Multibeam Pulsar Survey \citep{ckl+00}.  The pulsar is suitable for
the study of pulsar spin-down: it is one of the youngest known
pulsars, with an estimated age from timing of $1.7 \pm 0.1$ kyr, and
is notable for having one of the strongest surface magnetic strengths
of any known radio pulsar ($B \equiv 3.2 \times 10^{19}
(P\dot{P})^{1/2} \sim 4.1 \times 10^{13}$ G under the magnetic dipole
assumption). A measured second period derivative for the pulsar
\citep{ckl+00} can in principle be used in combination with a
constraint on $\alpha$ to test the Melatos model of pulsar spin-down
(see Section \ref{melatos-discussion}).

\section{Observations and Data Reduction}

We have analyzed polarimetry data taken with the Australia Telescope
Compact Array (ATCA; Frater, Brooks, \& Whiteoak 1992\nocite{fbw92})
and the Parkes radio telescope at center frequencies of 2496 and 1366
MHz, respectively. Details of the data analysis are presented below.

\subsection{ATCA 2496-MHz Data} 

Data were taken of PSR J1119$-$6127 with the ATCA using a 128-MHz
bandwidth centered on a frequency of 2496 MHz\footnote{Data were taken
simultaneously at 1384 MHz with the ATCA using the dual-band feed, but
Faraday smearing across the bandwidth from the large rotation measure
(see rotation measure estimate below) reduced the measured linear
polarization to less than 5\% of the intrinsic value.  We therefore do
not use the 1384-MHz ATCA data in the analysis here and do not mention
it further in this paper.} as part of a radio imaging campaign in
which supernova remnant SNR G292.2$-$0.5 was discovered; this is a
young remnant associated with the pulsar \citep{cgk+01a}. The ATCA
observations were conducted on 30 and 31 Oct 1998 in the 6D array
configuration using pulsar gating. The observing parameters, given in
Table \ref{tab:newparams}, are outlined with more extensive details of
the data analysis elsewhere \citep{c00, cgk+01a}.

Frequency channels contaminated with self-generated RFI were
automatically excised at the start of the analysis, and alternating
channels from the remaining set were preserved as a set of 13 channels
of width 8 MHz each, giving 104 MHz of usable bandwidth. Since there
is overlap between original adjacent channels, no sensitivity penalty
was incurred in the selection of alternate channels. After the data
were flagged and edited, the on-pulse data were selected from the
pulsar gating, and Stokes parameters were extracted for each frequency
channel at the pulsar's position. These were used to compute the
rotation measure (RM) for the pulsar.  Using the MIRIAD data analysis
package\footnote{See R. J. Sault \& N. E. B. Killeen, 1999, The MIRIAD
User's Guide (Sydney: Australia Telescope National Facility), found at
http://www.atnf.csiro.au/computing/software/miriad.}, Stokes $Q$ and
$U$ from each channel were converted into a PA $\psi$ according to:

\begin{equation} 
\psi = \frac{1}{2} \arctan \left( \frac{U}{Q} \right) .
\end{equation}

\begin{figure}
\epsscale{0.6}\plotone{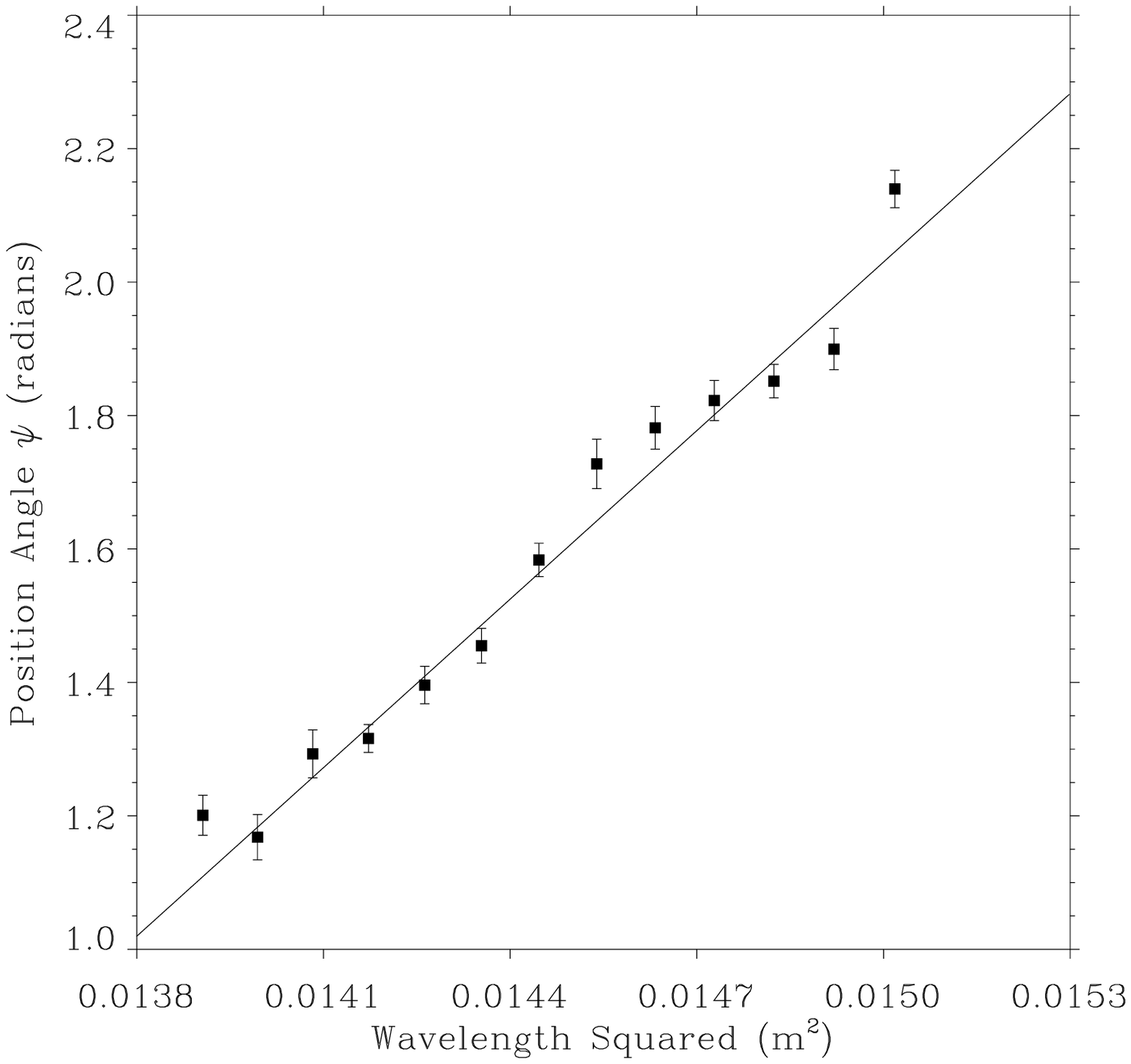} \figcaption[f1.eps]{Position angle as a
function of wavelength squared for the 2496-MHz pulsar-gated ATCA
data. 13 frequency channels of on-pulse data at the pulsar's position
were used.  The best fit line is overlaid, with a slope (RM) of $+842
\pm 23$ rad m$^{-2}$.
\label{fig:rm} }
\end{figure}

An uncertainty in each PA was also computed using this routine, and
the resulting PA for each channel was plotted against the square of
the wavelength for each channel (see Figure \ref{fig:rm}).  A linear
fit of the form $\psi = \psi_{0} + {\rm RM}\lambda^{2}$ was then
performed on the 13 data points to determine the RM. The best-fit
slope gave RM = $+842 \pm 23$ rad m$^{-2}$. The resulting Faraday
depolarization across the bandwidth at 2496 MHz was $\sim$ 4\%,
indicating that the polarization profile retains fidelity even in the
absence of a Faraday rotation correction.

The frequency channels were then summed, and Stokes parameters were
preserved for each of 32 pulse phase bins.  A mean off-pulse baseline
was subtracted from the total intensity profile, and the magnitude of
the linear polarization for each phase bin, $L = (Q^{2} +
U^{2})^{1/2}$, was computed and corrected for positive bias as
follows:

\begin{equation}
L = ( \vert L_{\rm obs}^2 - \langle L^2 \rangle_{\rm off} \vert
)^{1/2} .
\label{eqn:fraclin}
\end{equation}

\noindent $\langle L^2 \rangle_{\rm off}$ is the average value of the
square of $L$ for all off-pulse bins. Stokes $V$ represents the
circularly polarized intensity, with positive values corresponding to
left-circularly polarized radiation. The percentage of linear
polarization in the pulse profile, $\langle L \rangle / S$, was
computed as the mean fractional linear polarization for all on-pulse
bins, where $S$ is the flux from Stokes $I$.  The percentages of
circular and absolute circular polarization were likewise computed as
$\langle V \rangle / S$ and $\langle |V| \rangle / S$, respectively.

\subsection{Parkes 1366-MHz Data}

Data were taken of PSR J1119$-$6127 with the Parkes radio telescope on
16 and 18 Jan 1999. The observations were conducted using the
multibeam receiver \citep{s++96} and Caltech Correlator \citep{n94,
nms+97} at a center frequency of 1366 MHz covering a bandwidth of 128
MHz. Eight channels of width 16 MHz each were preserved.

Four separate integrations of 12 min each were summed, totaling 48
min. The four observations were taken in two sets of two consecutive
observations. The first set (taken on 16 Jan 1999) was separated by
two days from the second set (taken on 18 Jan 1999). Feed rotation
provided correction for parallactic angle variation during the
observations.  The observing parameters are presented in Table
\ref{tab:newparams}, and the observing technique was similar to the
one described by \citet{mhq98} and \citet{cmk01b}. 256 pulse phase
bins were preserved across the full profile, allowing a
high-resolution study of the behavior of the PA over the pulse phase.

An independent estimate of the RM using the Parkes data gave RM =
$+823 \pm 6$ rad m$^{-2}$ (see, e.g., Manchester, Han, \& Qiao
1998\nocite{mhq98}; Crawford, Manchester, \& Kaspi
2001b\nocite{cmk01b} for details on the technique of how the RM was
estimated). Although this RM is consistent with the ATCA RM estimate,
the very small uncertainty in the Parkes estimate cannot be readily
believed. Multibeam receiver instrumental effects were present which
affect the measured RM at a level significantly greater than the
quoted uncertainty, which is less than 1\% (see, e.g., Johnston
2002\nocite{j02} for details). We instead use the more reliable RM
estimate from the ATCA data.  However, a phase-rotation correction
could still be applied to the Parkes data prior to channel
summing. Trial phase rotations were applied until the resulting
measured linear polarization $L$ in the profile from the sum across
channels was maximized, indicating constructive addition of the linear
polarization (and the proper correction for Faraday
rotation). Uncorrectable Faraday smearing within the frequency
channels accounted for a reduction in the measured fractional linear
polarization of $\sim$ 4\%.

After phase-rotation and channel summing, the Stokes parameters were
used to determine polarization parameters for each bin in the same way
as for the ATCA data (described above).  The uncertainty in $\psi$ for
each profile bin was based on the scalar uncertainty determined from
the off-pulse rms of Stokes $I$. The linear polarization vector for a
given bin could deviate in any direction by this uncertainty, and the
corresponding deviation in the vector's PA was taken to be the
uncertainty in $\psi$. These uncertainties were used when computing
the RV best fit for the Parkes data.  33 data points in the profile
had PA uncertainty less than $15^{\circ}$, indicating significant and
measurable linear polarization. These points were used in the PA fit
(see Section \ref{rvfit}). Only the Parkes data were used for this fit
since the ATCA profile had insufficient resolution.

\section{Results and Discussion}

\subsection{Radio Polarization Properties of PSR J1119$-$6127}

\begin{figure}
\epsscale{0.4}\plotone{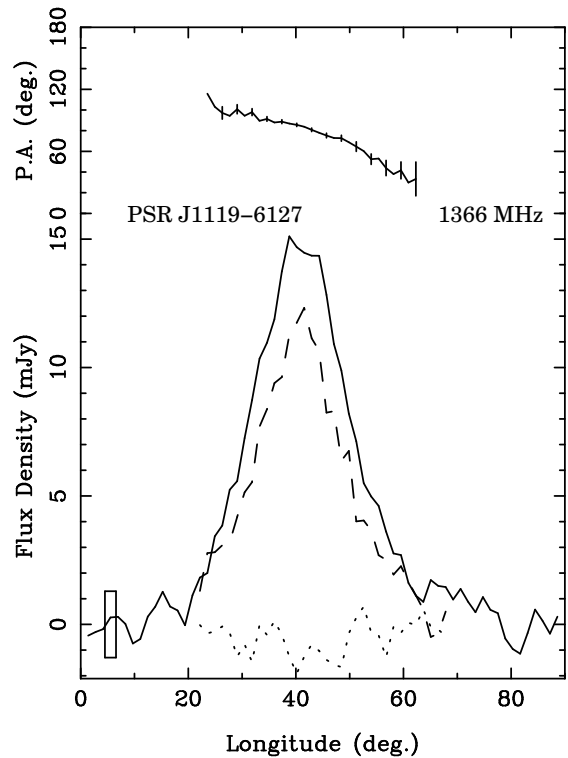} \figcaption[f2.eps]{1366-MHz
polarization profile of PSR J1119$-$6127 from the Parkes data.  The
full 360 degrees of phase of the profile spans 256 bins, but only the
portion in which the pulse appears is shown.  In the lower part of the
plot, the solid line indicates total intensity as a function of pulse
phase in degrees. The dashed and dotted lines indicate the linearly
and circularly polarized intensity, respectively. Positive values of
circular polarization correspond to left-circular polarization. The
height of the box in the lower left-hand corner is twice the baseline
scatter and does not reflect the additional uncertainty in the
measured polarization arising from multibeam receiver instrumental
effects. The upper part of the plot shows the PA plotted as a function
of pulse phase on the same axis. The pulsar is highly linearly
polarized, consistent with trends noticed for young energetic
pulsars. \label{fig:profile} }
\end{figure}

\begin{figure}
\epsscale{0.6}\plotone{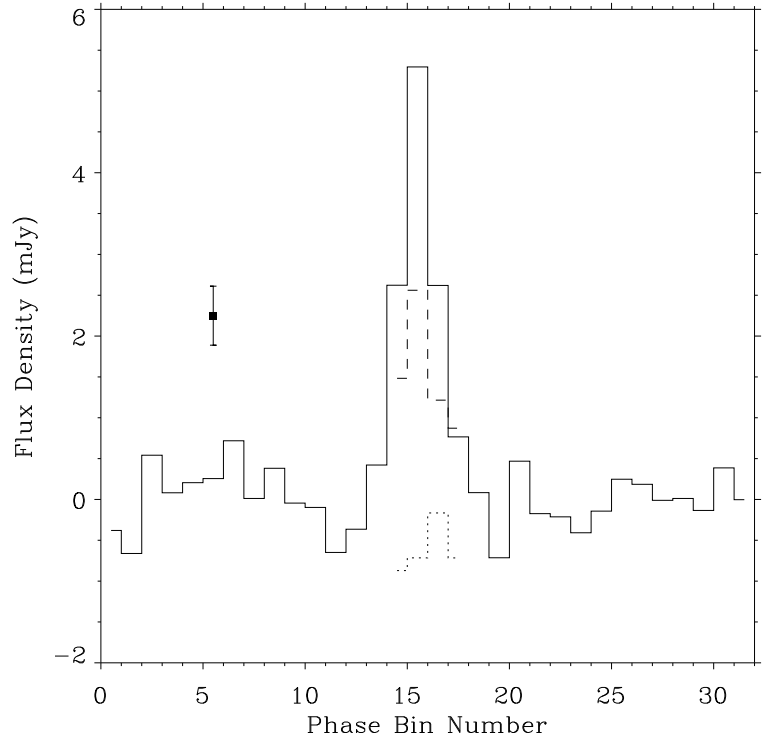} \figcaption[f3.eps]{2496-MHz
polarization profile of PSR J1119$-$6127 from pulsar-gated ATCA
data. The profile for one period (360 degrees of phase) is shown,
corresponding to 32 phase bins. The solid line indicates total
intensity while the dashed and dotted lines indicate the linearly and
circularly polarized intensity, respectively. Positive values of
circular polarization correspond to left-circular polarization. The
baseline scatter is indicated by the error bar to the left of the
profile.  The pulsar remains highly polarized at high radio
frequencies.
\label{fig:profile2} }
\end{figure}

Measured polarization parameters for PSR J1119$-$6127 from the
1366-MHz Parkes data and the 2496-MHz ATCA data are presented in Table
\ref{tab:newparams}, and the Parkes and ATCA polarization profiles are
shown in Figures \ref{fig:profile} and \ref{fig:profile2},
respectively. The on-pulse emission has strong linear polarization at
both frequencies, with a fractional linear polarization (scaled
upwards by 4\% in each case to correct for Faraday smearing across
finite bandwidths) of 77 $\pm$ 10\% and 56 $\pm$ 6\% at 1366 and 2496
MHz. The uncertainty in the linear polarization fraction measured in
the Parkes data includes the error introduced by multibeam receiver
polarization impurities. These impurities affect the measured circular
polarization and, to a lesser extent, the linear polarization
\citep{j02}.  It is clear that the pulsar remains highly polarized at
high radio frequencies. The circular polarization is weaker in both
cases.  The RM, measured using the 2496-MHz ATCA data, is $+842 \pm
23$ rad m$^{-2}$. The relation of the RM to the mean line-of-sight
interstellar magnetic field is given by \citep{mt77}:

\begin{equation}
\langle B_{\Vert} \rangle = 1.232 \frac{\rm RM}{\rm DM} \mu{\rm G}
\end{equation}

\noindent where DM is the dispersion measure in units of pc cm$^{-3}$
(DM = 707 pc cm$^{-3}$ for PSR J1119$-$6127). $\langle B_{\Vert}
\rangle$ for PSR J1119$-$6127 is $+1.47 \pm 0.04$ $\mu$G, where a
positive value corresponds to field lines pointing toward the
observer. This value is consistent with typical galactic magnetic
field strengths \citep{hmq99}.

The profile's high degree of linear polarization is particularly
noteworthy.  Another young radio pulsar, PSR J1513$-$5098, has similar
characteristics to PSR J1119$-$6127 (i.e., a $\sim$~2 kyr age, a very
large magnetic field, and a relatively long period for such a young
pulsar). \citet{cmk01b} report on radio polarization observations of
PSR J1513$-$5908 taken with Parkes at 1350 MHz and show that the
pulsar is essentially completely linearly polarized at this frequency.
At 1366 MHz, PSR J1119$-$6127 has a single, wide pulse, with a width
of $\sim 20^{\circ}$ as measured at 50\% of the peak, and a width of
$\sim 45^{\circ}$ as measured at 10\% of the peak.  This wide pulse is
again similar to the morphology of PSR J1513$-$5908, which at 1350 MHz
has pulse widths of $\sim 35^{\circ}$ and $\sim 95^{\circ}$ at 50\%
and 10\% of the peak, respectively. In general, the pulse morphology
of PSR J1119$-$6127 is similar to the single, wide, highly linearly
polarized profiles of the young pulsars presented by \citet{cmk01b}.

With its relatively large spin-down luminosity $\dot E \equiv 3.94
\times 10^{46} \dot{P}/P^{3} = 2.3 \times 10^{36}$ ergs s$^{-1}$, the
pulsar fits a positive trend noticed previously at 1400 MHz between
spin-down luminosity and degree of linear polarization (see, e.g.,
Figure 2 of \nocite{cmk01b}Crawford, Manchester, \& Kaspi 2001b).  The
pulsar also fits the association between small characteristic age (in
this case, $\tau_{c} \equiv P/2\dot{P} \sim 1.6$ kyr) and strong
linear polarization noticed by \citet{gl98}. Since spin-down
luminosity and characteristic age are correlated by definition
($\tau_{c} \sim 1/P^{2} \dot{E}$), this is not surprising.

\subsection{PA Swing and RV Fit from the Parkes Data}
\label{rvfit}

\begin{figure}
\epsscale{0.6}\plotone{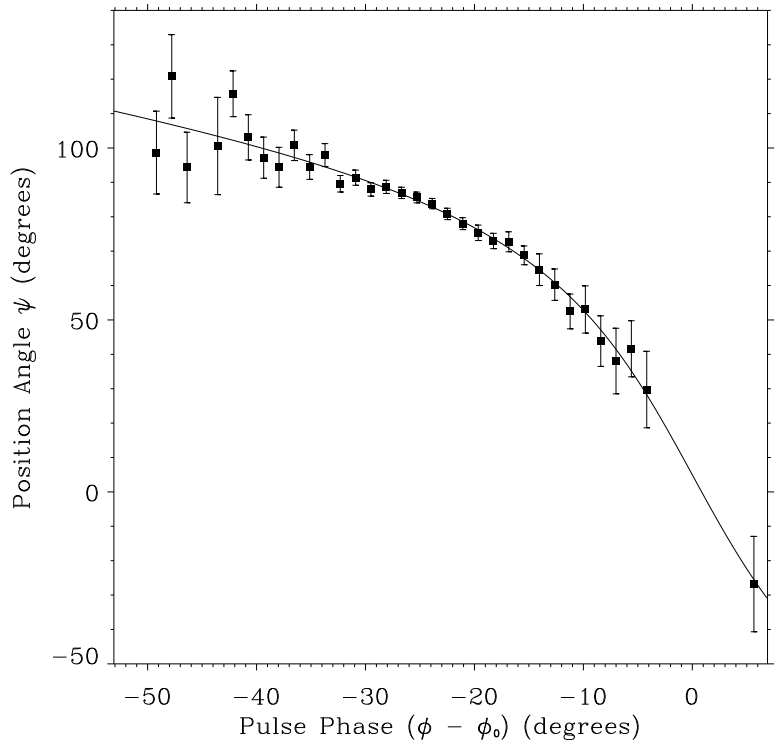} \figcaption[f4.eps]{PA as a function of
pulse phase for PSR J1119$-$6127 from the 1366-MHz Parkes data, with
the best-fit RV model overlaid. Data points with PA uncertainty less
than $15^{\circ}$ are shown and were used in the fit. The pulse phase
point $(\phi - \phi_0) = 0$, where the PA swing is greatest,
corresponds to the magnetic pole sweeping past the line of sight. The
peak of the profile, where the PA uncertainty is smallest, leads
$(\phi - \phi_0) = 0$, suggesting that the emission is emanating from
the leading edge of a conal beam. \label{fig:fit}}
\end{figure}

Measured PAs with uncertainty less than $15^{\circ}$ are shown as a
function of pulse phase in Figure \ref{fig:fit} for the 1366-MHz
Parkes data. Overlaid is the best-fit RV model. The fit used a
downhill simplex $\chi^2$-minimization algorithm in 4 dimensions
(e.g., \nocite{ptv+92}Press et al.\ 1992).  The model fits well, with
a best-fit $\chi^2$ of 16.5 with 29 degrees of freedom.  The
characteristic swing in PA is noticeable, with a maximum swing
occurring at $\phi_{0}$, as determined by the best fit.  There is
almost no radiation at the point $\phi = \phi_0$ and after; the pulse
peak leads the PA swing. This is consistent with the partial conal
beam structure interpretation set forth by \citet{lm88} and is similar
to the PA behavior seen for other young pulsars \citep{cmk01b}.
 
\begin{figure}
\epsscale{0.6}\plotone{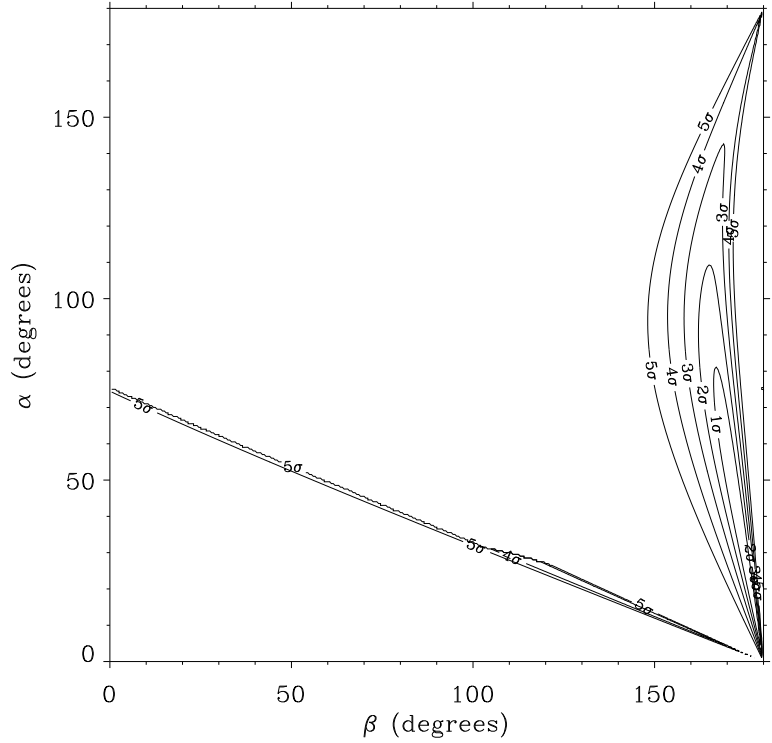} \figcaption[f5.eps]{Confidence regions
in $\alpha$ (magnetic inclination angle) and $\beta$ (impact
parameter) for the RV model best fit of the 1366-MHz Parkes PA
data. Contours at $1\sigma$, $2\sigma$, $3\sigma$, $4\sigma$, and
$5\sigma$ confidence levels are indicated. At the $3\sigma$ level,
$-20^{\circ} \approxlt \beta \approxlt 0^{\circ}$ and $\alpha
\approxlt 140^{\circ}$. This constraint on $\alpha$ is not sufficient
to test the spin-down model of Melatos (1997). \label{fig:contour}}
\end{figure}

Constraints from the RV fit on the parameters $\alpha$ and $\beta =
\zeta - \alpha$ are shown in Figure \ref{fig:contour}.  The
statistical constraints on the fit parameters imply $|\beta| \approxlt
20^{\circ}$ at the 3$\sigma$ confidence level.  However, $\alpha$ can
only be constrained to $\alpha \approxlt 140^{\circ}$ at the $3\sigma$
level owing to the limited data available. It is important to note
that while $\alpha$ and $\beta$ display little covariance, $\alpha$
and $\zeta$ have concomitantly great covariance.

\subsection{Testing the Melatos Model of Pulsar Spin-down}
\label{melatos-discussion}

\begin{figure}
\epsscale{0.6}\plotone{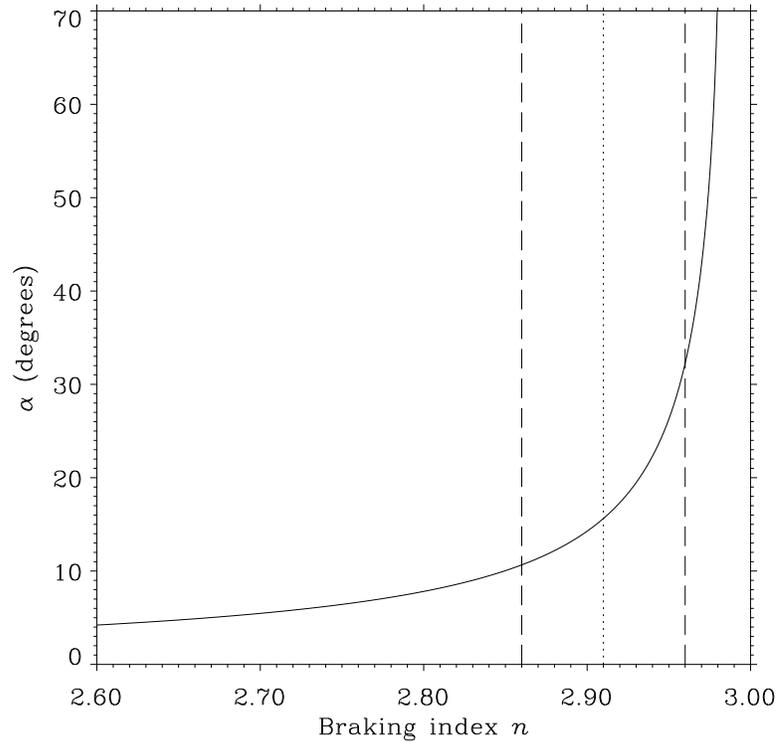} 
\figcaption[f6.eps]{Plot of magnetic inclination angle $\alpha$
as a function of braking index $n$ for PSR J1119$-$6127, as predicted
by the model of \citet{m97}.  The vertical lines correspond to the
range $n = 2.91 \pm 0.05$ measured for PSR J1119$-$6127
\citep{ckl+00}. This corresponds to $10^{\circ} \approxlt \alpha
\approxlt 32^{\circ}$ in the model. The greater slope around $n =
2.96$ indicates that reducing the uncertainty in $n$ would
significantly reduce the range of $\alpha$ that could be in agreement
with the model. \label{fig:model}}
\end{figure}

We have attempted to test the Melatos model of spin-down using an
estimate of $\alpha$ from the RV fit of the Parkes PA data (described
in Section \ref{rvfit}) and measurements of $n$ and $m$ from previous
timing observations.  \citet{ckl+00} measured a braking index $n =
2.91 \pm 0.05$ for PSR J1119$-$6127, which included uncertainty from a
glitch and timing noise. Within the Melatos model, this value of $n$
implies that $\alpha$ should lie between $10^{\circ}$ and $32^{\circ}$
(see Figure \ref{fig:model}).

A measurement of the second braking index $m$ is unavailable from the
current timing data; furthermore, possible future glitches make the
prospect of accurately measuring $m$ uncertain. If $m$ were to be
estimated from future timing observations, a significant test of the
spin-down model using $m$ would still require an independent geometry
constraint that excludes very small values of $\alpha$; for $\alpha$
less than a few degrees, the model's prediction of $m$ fluctuates
across a wide range of values.

Although the Parkes PA data are fit well by the RV model (see Section
\ref{rvfit}), the fit does not provide a meaningful statistical
constraint on $\alpha$. At the 3$\sigma$ level, $\alpha \approxlt
140^{\circ}$ in the fit. Since the model's prediction of the braking
index $n$ depends on the sine of $\alpha$, a useful constraint on $n$
requires a constraint on $\alpha$ better than $0^{\circ} < \alpha <
90^{\circ}$.  The $3\sigma$ constraint on $\alpha$ from these
observations is thus insufficient to test the Melatos model.
Additional PA data from future polarimetry observations may be able to
sufficiently constrain $\alpha$ and make a significant test of the
model possible. We estimate that with $\sim 12$ hr of polarization
observations at 1400 MHz with a similar system, we could obtain a
$3\sigma$ constraint on $\alpha$ that would be useful for the model
test.

As shown in Figure \ref{fig:model}, improvements in the measurement of
$n$ itself might also aid a test of this model. Reducing the
uncertainty in $n$ would reduce the range of observed $\alpha$ that
could be in agreement with the model.

\section{Conclusions} 

Using pulsar-gated 2496-MHz radio imaging data \citep{cgk+01a} taken
with the ATCA and archival 1366-MHz polarization data taken with the
Parkes telescope, we report on the polarization properties of PSR
J1119$-$6127, a pulsar notable for its youth and strong magnetic
field. A Faraday rotation measurement using the ATCA data gives a RM
of $+842 \pm 23$ rad m$^{-2}$ for the pulsar and a corresponding mean
line-of-sight magnetic field strength of $+1.47 \pm 0.04$ $\mu$G,
consistent with typical galactic magnetic field values.  The pulsar's
polarization profile shows a high degree of linear polarization ($\sim
75$\% at 1366 MHz and $\sim 55$\% at 2496 MHz), in agreement with
previously noticed trends for young pulsars at 1400 MHz \citep{cmk01b,
gl98}. The pulsar also has linear polarization and pulse morphology
characteristics which are similar to those seen for other young
pulsars (e.g., PSR J1513$-$5908).  A RV model fit of the observed PA
swing from the Parkes data constrains the impact parameter to $|\beta|
\approxlt 20^{\circ}$ and indicates that the pulse peak leads the PA
symmetry axis.  Additionally, the pulsar's profile consists of a
single wide component.  These features suggest emission from the
leading edge of a wide hollow cone beam, consistent with the partial
conal interpretation outlined by \citet{lm88}.

PSR J1119$-$6127's measurable braking index and clean polarization
profile suggest that it may be used in the future to test the model of
pulsar spin-down proposed by \citet{m97}. While constraints on the
magnetic inclination angle $\alpha$ obtained from a RV model fit to
the available Parkes PA data are inadequate for a significant test of
this model, further refinements from pulsar timing and additional
polarization observations could make such a test possible.

\acknowledgements

We thank Andrew Melatos and the referee Simon Johnston for insightful
comments and helpful suggestions for improving the manuscript.  We
also thank Bryan Gaensler for advice on the ATCA data analysis and
John Reynolds and John Sarkissian for providing the relevant Parkes
observing logs. N.C.K. was supported through a Research Experience
for Undergraduates supplemental research grant from the National
Science Foundation. The Parkes radio telescope and ATCA are part of
the Australia Telescope, which is funded by the Commonwealth of
Australia as a National Facility operated by CSIRO.

\clearpage

%\clearpage

%\clearpage

%\clearpage 

%\clearpage

%\clearpage

\begin{deluxetable}{lcc}
%\tabletypesize{\scriptsize}
\tablecaption{Observing Parameters and Measured Polarization Parameters for PSR J1119$-$6127. \label{tab:newparams}}
\tablewidth{0pt}
\tablehead{
\colhead{} & {} } 
\startdata
Telescope                             & Parkes    & ATCA      \\ 
Receiver                              & Multibeam & 13 cm     \\ 
On-source integration time (h)        & 0.8       & 9         \\
Center frequency (MHz)                & 1366      & 2496      \\
Bandwidth (MHz)                       & 128       & 104       \\
Number of frequency channels          & 8         & 13        \\
Number of bins in pulse profile       & 256       & 32        \\
Pulse width at 50\% of peak (deg)     & $\sim$ 20 & $\sim$ 15 \\
Pulse width at 10\% of peak (deg)     & $\sim$ 45 & $\sim$ 35 \\
$\langle L \rangle / S$ (\%)\tablenotemark{a}               & $77 \pm 10$\tablenotemark{d}   & $56 \pm 6$ \\
$\langle V \rangle / S$ (\%)\tablenotemark{b}               & $-8 \pm 15$\tablenotemark{d} & $-22 \pm 6$ \\
$\langle \vert V \vert \rangle / S$ (\%)\tablenotemark{c}   & $10 \pm 15$\tablenotemark{d}   & $22 \pm 6$ \\
                                      &           &              \\
Rotation measure, RM (rad m$^{-2}$)\tablenotemark{e}   & $+842 \pm 23$ &              \\
Mean line-of-sight magnetic field, $\langle B_\Vert \rangle$ ($\mu$G)\tablenotemark{f}  & $+1.47 \pm 0.04$ & \\
\enddata

\tablecomments{The ATCA observing parameters are also presented in
detail in \citet{cgk+01a}.}

\tablenotetext{a}{Fractional on-pulse linear polarization. Corrected for channel/bandwidth depolarization.}

\tablenotetext{b}{Fractional on-pulse circular polarization. Positive
values correspond to left circular polarization.}

\tablenotetext{c}{Fractional absolute on-pulse circular polarization.}

\tablenotetext{d}{Quoted uncertainty includes the contribution from
multibeam receiver instrumental effects (e.g., Johnston
2002\nocite{j02}).}

\tablenotetext{e}{Determined from 2496-MHz pulsar-gated ATCA data.}

\tablenotetext{f}{Positive values correspond to magnetic field lines
toward the observer.}

\end{deluxetable}

\end{document}